\documentclass[aip, amsmath,amssymb,reprint, longbibliography, final]{revtex4-2}
\usepackage[colorlinks,bookmarks=false,citecolor=blue,linkcolor=red,urlcolor=blue]{hyperref}

\usepackage[usenames,dvipsnames]{color}
\usepackage{soul}

\usepackage{amsfonts}
\usepackage{amsmath}

\usepackage{graphicx}% Include figure files
\usepackage{dcolumn}% Align table columns on decimal point
\usepackage{bm}% bold math
\usepackage[utf8]{inputenc}
\usepackage[T1]{fontenc}
\usepackage{mathptmx}

\usepackage[dvipsnames,table]{xcolor}

\usepackage{siunitx}
\mathchardef\mhyphen="2D
\usepackage{floatrow}
\usepackage{multirow}
\usepackage{upgreek}
\usepackage{makecell}
\usepackage{tabularx}
\usepackage{physics}

\newcommand{\suppl}[1]{Supplementary}
\newcommand{\eqnref}[1]{Eq.~\eqref{eq:#1}}

\newcommand{\tabref}[1]{Tab.~\ref{tab:#1}}

\newcommand{\figref}[1]{Fig.~\ref{fig:#1}}

\newcommand{\figrefL}[1]{Figure~\ref{fig:#1}}

\newcommand{\subfigref}[2]{Fig.~\hyperref[fig:#1]{\ref*{fig:#1}#2}}
\newcommand{\Supplsubfigref}[2]{\suppl{}~Fig.~\hyperref[fig:#1]{\ref*{fig:#1}#2}}
\newcommand{\SupplsubfigrefL}[2]{\suppl{}~Figure~\hyperref[fig:#1]{\ref*{fig:#1}#2}}
\newcommand{\SupplsubfigrefS}[2]{Fig.~\hyperref[fig:#1]{\ref*{fig:#1}#2}}
\newcommand{\subfigrefL}[2]{Figure~\hyperref[fig:#1]{\ref*{fig:#1}#2}}
\newcommand{\subfigsref}[3]{Figs.~\hyperref[fig:#1]{\ref*{fig:#1}#2}-\hyperref[fig:#1]{\ref*{fig:#1}#3}}
\newcommand{\Supplsubfigsref}[3]{\suppl{}~Figs.~\hyperref[fig:#1]{\ref*{fig:#1}#2}-\hyperref[fig:#1]{\ref*{fig:#1}#3}}
\newcommand{\subfigsrefL}[3]{Figures~\hyperref[fig:#1]{\ref*{fig:#1}#2}-\hyperref[fig:#1]{\ref*{fig:#1}#3}}
\newcommand*{\figlab}[1]{\textbf{#1},}

\newcommand*{\smallR}[1]{$R_0 = \SI{364}{\nano\meter}$}
\newcommand*{\largeR}[1]{$R_0 = \SI{648}{\nano\meter}$}

\newcommand*{\fp}[1]{$F_{\mathrm{P}}$}
\newcommand*{\lowT}[1]{$T = \SI{4.2}{\kelvin}$}
\newcommand*{\methods}[1]{Methods}

\DeclareSIUnit\ML{ML}
\DeclareSIUnit\px{px}
\newcommand*{\SM}[1]{\suppl{}~Information}

\usepackage{array}
\usepackage{tabularx}
\usepackage{arraycols}

\usepackage[flushleft]{threeparttable}

\usepackage{booktabs}

\usepackage{etoolbox}
\robustify\bfseries
\robustify\itshape

\newcolumntype{E}{>{\centering\arraybackslash}X}
\newcommand{\ra}[1]{\renewcommand{\arraystretch}{#1}}

\usepackage{microtype}
\usepackage{rotating}
\usepackage{textcomp}

\newcommand*{\oldtext}[2][Old text: ]{\textit{#1}{\textcolor{Red}{#2}}}

\DeclareMathOperator{\arcsinh}{arcsinh}

\makeatletter
\let\@fnsymbol\@fnsymbol@latex
\@booleanfalse\altaffilletter@sw
\makeatother

\begin{document}
\preprint{AIP/123-QED}

\newcommand*{\papertitle}{Efficient passivation of III-As(P) photonic interfaces}

\author{Yury~Berdnikov}
\affiliation{DTU Electro, Department of Electrical and Photonics Engineering, Technical University of Denmark, Ørsteds Plads 343, DK-2800 Kongens Lyngby, Denmark}

\author{Pawe\l{}~Holewa}
\affiliation{DTU Electro, Department of Electrical and Photonics Engineering, Technical University of Denmark, Ørsteds Plads 343, DK-2800 Kongens Lyngby, Denmark}
\affiliation{Department of Experimental Physics, Faculty of Fundamental Problems of Technology, Wroc\l{}aw University of Science and Technology, Wyb. Wyspia\'{n}skiego 27, 50-370 Wroc\l{}aw, Poland}
\affiliation{NanoPhoton - Center for Nanophotonics, Technical University of Denmark, Ørsteds Plads 345A, DK-2800 Kongens Lyngby, Denmark}

\author{Aurimas~Sakanas}
\affiliation{DTU Electro, Department of Electrical and Photonics Engineering, Technical University of Denmark, Ørsteds Plads 343, DK-2800 Kongens Lyngby, Denmark}

\author{Jan Mikołaj Śmigiel}
\affiliation{Department of Experimental Physics, Faculty of Fundamental Problems of Technology, Wroc\l{}aw University of Science and Technology, Wyb. Wyspia\'{n}skiego 27, 50-370 Wroc\l{}aw, Poland}

\author{Pawe\l{}~Mrowi\'{n}ski}
\affiliation{Department of Experimental Physics, Faculty of Fundamental Problems of Technology, Wroc\l{}aw University of Science and Technology, Wyb. Wyspia\'{n}skiego 27, 50-370 Wroc\l{}aw, Poland}

\author{Emilia~Zi\k{e}ba-Ost\'{o}j}
\affiliation{Department of Experimental Physics, Faculty of Fundamental Problems of Technology, Wroc\l{}aw University of Science and Technology, Wyb. Wyspia\'{n}skiego 27, 50-370 Wroc\l{}aw, Poland}

\author{Kresten~Yvind}
\affiliation{DTU Electro, Department of Electrical and Photonics Engineering, Technical University of Denmark, Ørsteds Plads 343, DK-2800 Kongens Lyngby, Denmark}
\affiliation{NanoPhoton - Center for Nanophotonics, Technical University of Denmark, Ørsteds Plads 345A, DK-2800 Kongens Lyngby, Denmark}

\author{Alexander~Huck}
\affiliation{Center for Macroscopic Quantum States (bigQ), Department of Physics, Technical University of Denmark, DK-2800 Kongens Lyngby, Denmark}

\author{Marcin~Syperek}
\affiliation{Department of Experimental Physics, Faculty of Fundamental Problems of Technology, Wroc\l{}aw University of Science and Technology, Wyb. Wyspia\'{n}skiego 27, 50-370 Wroc\l{}aw, Poland}

\author{Elizaveta~Semenova}
\email{esem@fotonik.dtu.dk}
\affiliation{DTU Electro, Department of Electrical and Photonics Engineering, Technical University of Denmark, Ørsteds Plads 343, DK-2800 Kongens Lyngby, Denmark}
\affiliation{NanoPhoton - Center for Nanophotonics, Technical University of Denmark, Ørsteds Plads 345A, DK-2800 Kongens Lyngby, Denmark}


\begin{abstract}
Surface effects can significantly impact the performance of nanophotonic and quantum photonic devices, especially as the device dimensions are reduced.
In this work, we propose and investigate a novel approach to surface passivation to mitigate these challenges in photonic nanostructures with III-As(P) quantum wells defined by a dry etching process.
The nanostructures are annealed under the phosphine (PH$_3$) ambient inside a metal-organic vapor phase epitaxy chamber to eliminate surface and subsurface defects induced during the dry etching and subsequent oxidation of the etched sidewalls.
Moreover, encapsulation of the active material with a wider bandgap material allows for maintaining the band structure of the device, mitigating band bending effects.
Our findings reveal an almost order of magnitude reduction in the surface recombination velocity from $\SI{2E3}{\centi\meter\per\second}$ for the PH$_3$ annealing compared to $\SI{1.5E4}{\centi\meter\per\second}$ for the non-passivated structures and $\SI{5E3}{\centi\meter\per\second}$ for the standard method based on (NH$_4$)$_2$S wet treatment followed by Al$_2$O$_3$ encapsulation.
A further reduction to $\SI{5E2}{\centi\meter\per\second}$ is achieved for the InP-regrown samples.
Additionally, we develop a model accounting for the impact of surface charges in the analysis of time-resolved photoluminescence curves and demonstrate that the proposed passivation method effectively reduces the surface charge density on the sidewalls of the studied quantum well-based photonic nanostructures.
\end{abstract}

\title{\papertitle{}}

\maketitle

\section*{Introduction}

\begin{figure*}[tb]
  \centering
  \includegraphics[width=\textwidth]{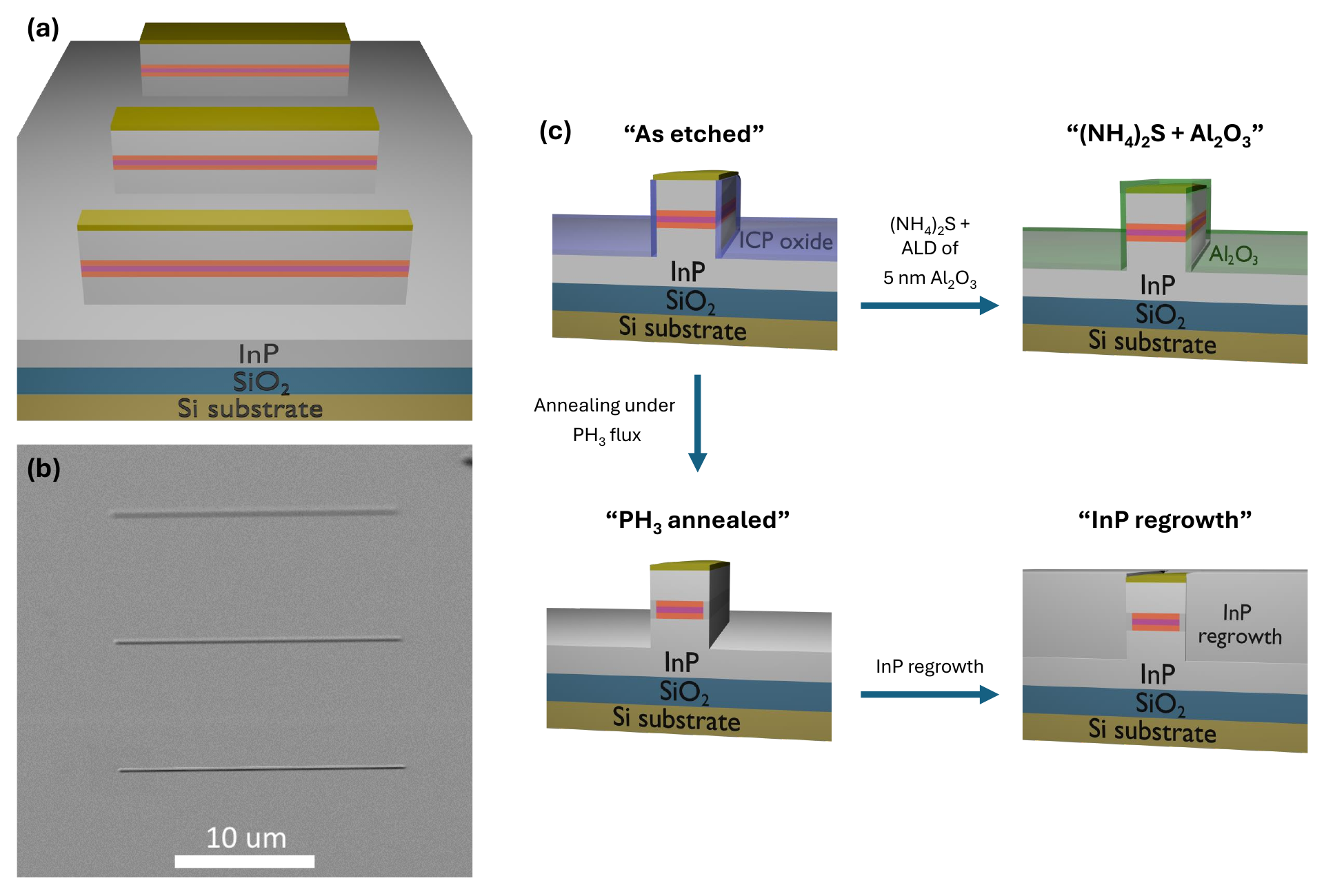} 
  \caption{\figlab{a}~Scheme of the fabricated nanoridges with III-V QWs (red). 
  \figlab{b}~An exemplary top view SEM image of the etched nanoridges.
  \figlab{c}~Schematic of the applied passivation process.}
  \label{fig:1}
\end{figure*}

The advancement of nanophotonics and quantum photonics has been a key driving force behind the development of nanofabrication techniques, facilitating the continued miniaturization of photonic devices, ultimately enabling the coherent control of light even at sub-wavelength scales~\cite{Moerk}.
Consequently, this reduction in the size of photonic components significantly increases the surface-to-volume ratio, thus increasing the impact of the surface on the device performance. 
Generally, semiconductor surfaces can have a detrimental effect on the performance of photonic devices, caused by non-radiative carrier recombination through surface states in the bandgap of the optically active region~\cite{coldren2012book,laukkanen2024bridging} or charge noise~\cite{Manna2020}.
In III-V materials that contain either As or P as the group V element -- referred to as III-As(P) -- and emit in the near-infrared range, surface defects are typically caused by uncontrolled surface oxidation.
This oxidation leads to the formation of gap states associated with As dangling bonds and As-As dimers~\cite{robertson2015defect}.
In addition to oxidation, ion-induced damage during dry etching of III-V materials, a process broadly used for device fabrication, could be a source of various types of surface and crystal point defects in volume~\cite{seong2020surface, djie2004experimental}.

Subsequently, considerable efforts from research groups worldwide have focused on surface state passivation to mitigate non-radiative losses and preserve the optical and quantum efficiency of miniaturized devices.
Several methods to reduce the detrimental effect of surface-related defects in III-V nanostructures have been reported so far. 
They include, first of all, native oxide removal steps.
The most widely used approach is the wet-chemical treatment of surfaces with ammonium or sodium sulfides~\cite{oigawa1991universal, amarnath2005electrically}. 
Alternatively to sulfur-based chemicals, surface nitridation was also used to reduce non-radiative recombination in GaAs nanostructures~\cite{alekseev2015nitride, guha2017surface}.
These wet-chemical approaches are usually combined with atomic layer deposition (ALD) and chemical vapor deposition (CVD) of amorphous layers to prevent surface re-oxidation~\cite{higuera2017ultralow, crosnier2015subduing}.
Other works also report the possibility of oxide removal from III-As surfaces inside the ALD chamber (termed "self-cleaning") before oxide or nitride deposition~\cite{hinkle2008gaas, timm2018self}.
The above methods aim to reduce the oxidation-related defects but do not treat the surface and sub-surface crystal defects caused by bombardment with ions during the dry etch.
Therefore, an additional etch in diluted H$_2$SO$_4$:H$_2$O$_2$:H$_2$O solution is used between the ion etching and surface passivation~\cite{boroditsky2000surface, crosnier2015subduing}.
This approach implies wet etch removal of an outer layer up to tens of nanometers thickness of III-V materials, which could be incompatible with shape imperfection-sensitive structures such as photonic crystals, microresonators, and cavities with extreme dielectric confinement~\cite{Xiong2024}.

Alternatively, structural defects can be at least partially healed during high-temperature annealing.
However, at high temperatures, the desorption of V$^{\rm th}$ group atoms occurs, which can be compensated for by supplying the corresponding adatom precursors~\cite{yoon1999effects, Berdnikov2024}.
Besides, annealing III-V materials with V$^{\rm th}$ group overpressure at elevated temperature is a preliminary oxide removal step in molecular beam epitaxy (MBE) and metal-organic vapor phase epitaxy (MOVPE).
Moreover, within this approach, near-surface As atoms can be replaced with other V$^{\rm th}$ group atoms, for example, P atoms.
The resulting P-reach surface layer has a bandgap wider than that of the As-rich part of the nanostructure and, therefore, separates the carrier confinement region of a photonic structure from the surface.
This process also referred to previously as "phosphidization", has been shown to effectively passivate the surfaces of GaAs layers~\cite{sugino1999phosphidization} and nanowires~\cite{haggren2014strong}.

In this study, we systematically investigate the surface state passivation in photonic nanostructures containing III-As(P) surfaces formed via dry etching by annealing them in phosphine ambient within the MOVPE chamber.
This method leverages the previously described replacement of near-surface arsenic atoms with phosphorus to achieve the required surface and bandgap modification.
We study InP/InGaAlAs/InGaAsP quantum well (QW) structures fabricated into ridges and compare the impact of our surface passivation approach with two other methods to treat the structures: the first one is the widely used approach to surface passivation using (NH$_4$)$_2$S wet treatment and the second one is complete regrowth with InP which leads to the formation of buried heterostructure~\cite{sakanas2019bh_thesis, Dimopoulos2022}.
We systematically investigate the impact of surface passivation on the carrier recombination in the InP ridges with embedded strain-compensated QWs and with the dry-etched sidewalls.
The sample fabrication steps are typical for nano- and quantum photonic device fabrication, and the studied nanoridges with QWs can be considered a model structure for device active regions. 

\section*{Methods and materials}

To investigate the surface recombination dynamics in the structures with properties as similar as possible to nano- and quantum photonic devices, we follow the main processing steps of QWs/InP on Si device fabrication~\cite{sakanas2019bh_thesis, Dimopoulos2022}.     
The InP layers with sandwiched In$_{0.78}$Ga$_{0.22}$As$_{0.85}$P$_{0.15}$/In$_{0.46}$Al$_{0.29}$Ga$_{0.25}$As QWs and In$_{0.47}$Ga$_{0.53}$As sacrificial etch-stop layer were grown on a 2" InP (001)-oriented wafer by MOVPE in a low-pressure TurboDisc$^{\textregistered}$ D125 reactor.  
Hydrogen (H$_2$) was used as the carrier gas, trimethylindium (TMIn), trimethylgalium (TMGa), and trimethylaluminum (TMAl) as the III$^{\rm rd}$ group precursors, while phosphine (PH$_3$), tertiarybutylphosphine (TBP), and arsine (AsH$_3$) were used as precursors of the V$^{\rm th}$ group.
The III-V structure was directly bonded to the central area of a 4" Si wafer covered by $\SI{1100}{\nano\meter}$ thermal oxide. 
Then, the InP substrate was removed by wet etching in HCl, which terminated at the InGaAs etch-stop layer.
The latter layer was subsequently removed in H$_2$SO$_4$:H$_2$O$_2$:H$_2$O solution.

In a broad range of photonic structures, the etched sidewalls, where the active area is exposed to the environment, play a crucial role in unwanted carrier recombination.
To access the sidewall recombination, we fabricate and investigate the arrays of ridges with varied widths etched in the III-V on Si wafer.
The comparison of nanoridges of different widths allows us to decouple the surface and bulk contributions to carrier recombination~\cite{boroditsky2000surface, andrade2021sub}.
The ridge patterns were defined by electron beam lithography using HSQ negative tone resist.
After exposure and development, we used the HBr-based dry etch process by inductively coupled plasma (ICP RIE) to define III-V ridges, leaving the QWs only under the masked areas. 
\subfigrefL{1}{a} schematically illustrates the obtained arrays of ridges with a length of $\SI{30}{\micro\meter}$ and the width varying between $\SI{200}{\nano\meter}$ and $\SI{3500}{\nano\meter}$. 

After the nanoridge fabrication, the wafer was cleaved to $\SI{5}{\milli\meter}\times\SI{5}{\milli\meter}$ chips.
The samples at this fabrication stage are referred to "as etched" (no further treatment).
Different chips originating from the same wafer were used for each surface treatment approach, as summarized in \subfigref{1}{c}.
In our approach to surface passivation, the etched structures were first wet-chemically treated in NH$_4$OH aqueous solution for $\SI{8}{\min}$ to remove the native oxide layer formed on the sidewalls after the ICP etch.
The chips were then transferred to the MOVPE chamber for annealing under PH$_3$ flux at $\SI{600}{\degreeCelsius}$ for $\SI{10}{\min}$.
After this step, some of the samples were taken out from the MOVPE chamber, and we refer to these samples as "PH$_3$ annealed". 
The samples remaining in the MOVPE reactor were then overgrown with an InP layer, and we refer to these samples as "InP regrowth".
As an alternative method, some samples obtained after the ICP etching were treated for $\SI{10}{\min}$ in $\SI{20}{\percent}$ aqueous solution of (NH$_4$)$_2$S at room temperature and then immediately transferred to a chamber for ALD, where the structures were covered with a $\SI{10}{\nano\meter}$ thick layer of Al$_2$O$_3$.
We refer to the samples obtained after this rather commonly used approach as "(NH$_4$)$_2$S + Al$_2$O$_3$".

To evaluate the carrier recombination dynamics in the fabricated structures and the impact of the ridge processing, we employed the time-resolved photoluminescence (TRPL) technique.
TRPL measurements were carried out using superconducting single-photon detectors with a time-correlated single-photon counting (TCSPC) module.
Carriers were optically excited using a pulsed laser with a central wavelength of $\SI{800}{\nano\meter}$ and a pulse duration of $\SI{500}{\pico\second}$.
The same objective lens with $20\times$ magnification was used to focus the excitation laser and collect the photoluminescence signal.

\section*{Results and discussion }

The ridge surface state density is modified by the processing and acts as non-radiative recombination centers for photo-generated carriers imprinting the photoluminescence decay and thus probing the interface quality.
To separate the effects originating from the bulk and the surface, we compare the experimental results obtained from ridges of various widths\cite{boroditsky2000surface,andrade2021sub}.
With this approach, we can estimate the surface recombination velocity $s$ based on the total recombination rate $1/\tau$ obtained from the TRPL measurements and expressed as the sum of contributions from the bulk and surface recombination: 

\begin{equation}
    \frac{1}{\tau} = \frac{1}{\tau_{\rm bulk}} + \frac{2 s}{w},
    \label{eq:1}
\end{equation}
where the first term on the right-hand side, $1⁄ \tau_{\rm bulk}$, is the carrier recombination rate in bulk, and the surface recombination rate $2s/w$ scales with the ridge width $w$.  

\subsection*{Stretched exponential analysis}
\begin{figure*}[htb]
  \centering
  \includegraphics[width=0.9\textwidth]{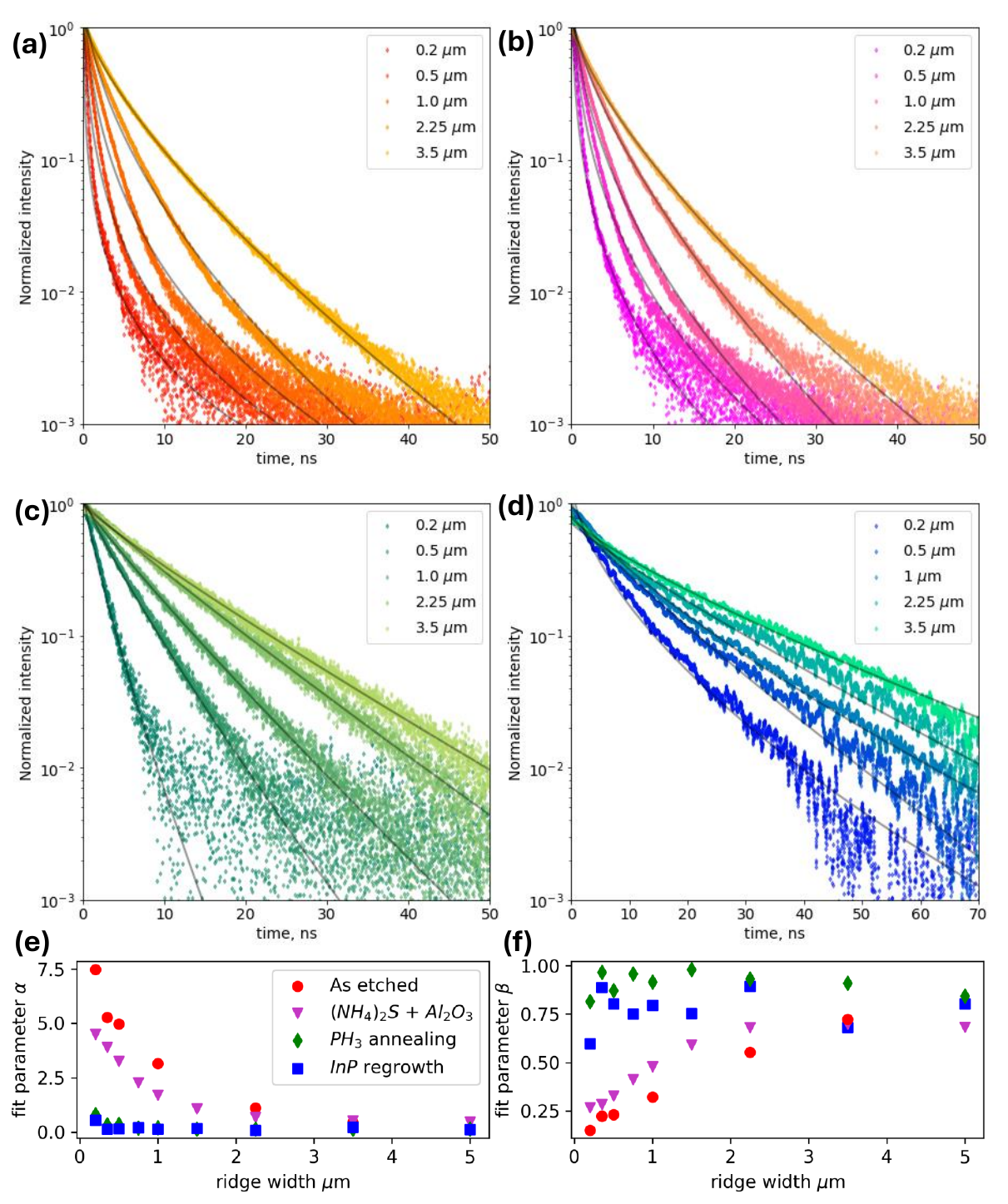} 
  \caption{Analysis of the TRPL decays for ridges with width from $\SI{0.2}{\micro\meter}$ to $\SI{3.5}{\micro\meter}$ and different surface treatment.
  a--d, TRPL decay curves for \figlab{a}~as etched, \figlab{b}~ (NH$_4$)$_2$S + Al$_2$O$_3$ treated, \figlab{c}~PH$_3$ annealed, \figlab{d}~InP regrown samples.
  e--f, Analysis of \figlab{e}~decay rate parameter $\alpha$, \figlab{f}~temporal stretching parameter $\beta$ [see \eqnref{2}].}
  \label{fig:2}
\end{figure*}

\figrefL{2} shows the room-temperature TRPL traces for the processed structures, plotted as a function of the ridge width spanning the range of $\SIrange{200}{3500}{\nano\meter}$.
The traces are normalized to maximum intensity and shown in log-linear coordinates for clarity of comparison. 
Panels (a)-(d) in \figref{2} correspond to different surface treatments as illustrated in \subfigref{1}{c}. 
The measured TRPL curves generally have a non-exponential shape, which is more pronounced in the cases of "As etched" and "PH$_3$ annealed" structures shown in \subfigsref{2}{a}{b}, respectively.
The non-exponential shape of the TRPL traces indicates the contribution of different radiative and non-radiative relaxation channels in the carrier recombination process~\cite{xing2017novel, andrade2021sub}.
In this case, the recorded TRPL histograms can be approximated with a stretched exponential function~\cite{johnston2006stretched}:

\begin{equation}
    I = I_0  \exp\left(- \alpha t^\beta\right),
    \label{eq:2}
\end{equation}
where $I_0$ is the maximum intensity, $\alpha$ is the decay rate, and $\beta$ is a temporal stretching parameter.

For the case of $\beta = 1$, the carrier lifetime is time-independent, and the recombination rate $1⁄\tau$ in \eqnref{1} is identical to $\alpha$.
For the case of $\beta < 1$, the carrier lifetime increases over time, and the use of \eqnref{1} requires further discussion. 

The values of $\alpha$ and $\beta$ extracted from the fits to TRPL decays are summarized in \subfigsref{2}{e}{f}.
For all samples, we observe an increase of $\alpha$ with decreasing width of the ridges, which we attribute to an increased contribution of non-radiative recombination channels at the sidewalls as the surface-to-volume ratio increases for narrower structures.

\subfigrefL{2}{f} shows that in the cases of InP regrowth and passivation by annealing in phosphine, the obtained values of $\beta$ vary only slightly with the ridge width $w$ and stay just below unity.
This observation witnesses slow variation of recombination rate (and thus lifetime) throughout the carriers relaxation without strong impact of the ridge widths.
In contrast, non-passivated samples and those passivated with (NH$_4$)$_2$S + Al$_2$O$_3$ treatment show a decrease of $\beta$ towards zero as a function of decreasing ridge width $w$.
The observed decrease of $\beta$ in these samples suggests the increase of the carrier lifetime throughout the carrier relaxation process.
This is more pronounced in narrow ridges with higher surface-to-volume ratios and, thus, a stronger impact of surface-related effects.

The stretched exponent model is helpful for the quantitative comparison of TRPL data systematically obtained for samples with various surface treatments and widths.
Still, it does not provide detailed information about the intrinsic carrier dynamics that results in the observed relaxation curves.
Therefore, further considerations beyond the widely used model of \eqnref{1} are needed to interpret the measured TRPL decays and observed lifetime increase during the carrier relaxation process. 

To interpret the observed dependencies in TRPL curves, we first employ an approach based on the steady-state solution of the rate equation for the average density of photo-generated carriers $N(x,t)$ in the QW inside the ridges~\cite{andrade2021sub, xing2017novel}.
Here, the average density $N$ refers to $N^2 = pn$ and includes densities of electrons $n$ and holes $p$.
We consider the high injection approximation, where the density of photo-generated carriers far from the sidewall surface $N$ is much larger than the equilibrium value $N_i = n_i p_i$.
With these approximations, the radiative recombination can be written as a quadratic function of $N$:

\begin{equation}
    I = B N^2,
    \label{eq:3}
\end{equation}
where the parameter $B$ characterizes the radiative recombination rate, and $I$ is the measured PL signal intensity.
For the case of a TRPL measurement when carriers are generated instantaneously, the rate equation can be written in the form: 

\begin{equation}
    \frac{\dd{N}}{\dd{t}} = - A N - B N^2,	
    \label{eq:4}
\end{equation}
where the term $A N = R_{\rm surf}$ is the simplified expression for the rate of non-radiative recombination and with $N_0 = N(t=0)$.

Regarding the constant $A$, following the approach of Ref.~\onlinecite{andrade2021sub}, one can solve \eqnref{4} for $N(t)$ and use it in \eqnref{3} to obtain the model expression for the intensity $I(t)$.
Analyzing the limits of \eqnref{4}, one can observe two types of dependencies:
(i) for the case when non-radiative recombination dominates, $A \gg B N(t=0)$, the relaxation is purely exponential, $I(t)\sim \exp(-2At)$  and implies $1/\tau = 2A $, and (ii) for the case when radiative recombination dominates, $A \ll BN_0$, and the relaxation curve can be described by $I(t) \sim 1/(1+BN_0 t)^2$. 
Hence, we relate the non-exponential character ($\beta \approx 0.8$) of the measured TRPL traces from wide ridges and the ridges with InP regrowth to the considerable contribution of radiative recombination relative to a reduced contribution of the surface recombination.
Meanwhile, surface recombination is expected to dominate in the case of narrow ridges in non-passivated and (NH$_4$)$_2$S + Al$_2$O$_3$-passivated samples.
Therefore, the observed decrease in $\beta$ cannot be related to the contribution of the radiative channel.
Thus, explaining the observed non-exponential shape of the TRPL curves in this case requires more detailed considerations. 

\subsection*{Surface recombination rate}
To add the surface-related effects into the analysis, we first consider the Shockley-Read-Hall model for the recombination rate $R_{\rm surf}$ applied to the case of sidewall surfaces\cite{shockley1952statistics, hall1952electron}:

\begin{equation}
    R_{\rm surf}=\frac{p_s n_s-n_i^2}{S_p^{-1} (n_s + n_t)  + S_n^{-1}  (p_s + p_t)},
    \label{eq:5}
\end{equation}
where $p_s$ and $n_s$ are hole and electron densities in the vicinity of the sidewall surfaces, $p_t$  and $n_t$ are corresponding trap densities, and $S_p$ and $S_n$ are the corresponding surface recombination rates.
We assume $S_p = S_n = S_0$ for simplicity in the following. 
We also assume the validity of the quasi-Fermi level approximation, which implies exponential dependencies for $p_s = p_b \exp(v_s)$ and $n_s = n_b \exp(-v_s)$ with 
$v_s = eV_s/kT$ and $n_b$ ($p_b$) denoting the electron (hole) concentration far from the surface.
To estimate $n_s$ and $v_s$, we follow the well-known approach based on the balance between the surface charge density $Q_s$ and the charge density within the space-charge regions $Q_{sc} = -Q_s$~\cite{monch2013semiconductor, luth2013surfaces}.
The charge density in the space-charge region $Q_{sc}$ can be expressed as $Q_{sc} = \epsilon \epsilon_0 E_s$ in terms of the surface field $E_s$. 
By definition, the surface field is the derivative of the potential $E=-\dd{V}/\dd{x}$. Therefore, it can be found from the surface potential $v=eV/kT$ within the space-charge regions, which is determined by the Poisson equation~\cite{monch2013semiconductor, luth2013surfaces}:

\begin{equation}
    \frac{\dd[2]{v}(x, n_b, Q_s)}{\dd{x^2}} = -\frac{e^2}{\epsilon \epsilon_0 kT} \left( n_b  (e^{-v} - 1) - p_b (e^{v} - 1) \right),	
    \label{eq:6}
\end{equation}
where $x$ is the distance from the sidewall surface. 
Then, considering the case of high injection levels with $n_b,p_b \gg n_b^0 ,p_b^0 \gg n_t,p_t$, integration of \eqnref{6} over $V$ results in 
$Q_s \approx 2 \sqrt{ 2 \epsilon kT p_b}  \sinh(v_s/2)$ (more details of calculations are given in Appendix~\ref{SI:Band-bending}).
We assume the surface defects to be fully ionized, and thus $Q_s= e N_s = $ \textit{const} does not change with the carrier concentration, where $N_s$ is the surface charge density.
With these considerations, \eqnref{5} for the surface recombination simplifies to:

\begin{equation}
    R_{\rm surf} \approx \frac{1}{2} \frac{S_0  n_b}{1+c/p_b }, 
    \label{eq:7}
\end{equation}
with $c = Q_s^2/4N_0  \epsilon \epsilon_0 kT$.
More detailed derivation is given in Appendix~\ref{SI:Surface-recombination-rate}.

Therefore, based on \eqnref{7}, one can expect the contribution of surface recombination to follow $R_{\rm surf} \sim p_b  n_b = N^2$ for large band bending when $V_s \gg kT$.
In this case, the surface recombination rate is quadratic as a function of charge density in contrast to  $R_{\rm surf} \sim N$ previously considered in \eqnref{4}.
As discussed above, the non-radiative recombination rate changes non-linearly with the carrier density.
It implies a non-exponential decay of the TRPL curves as we observe for low-intensity levels in \subfigsref{2}{a}{b}.   

\begin{figure*}[htb]
  \centering
  \includegraphics[width=0.9\textwidth]{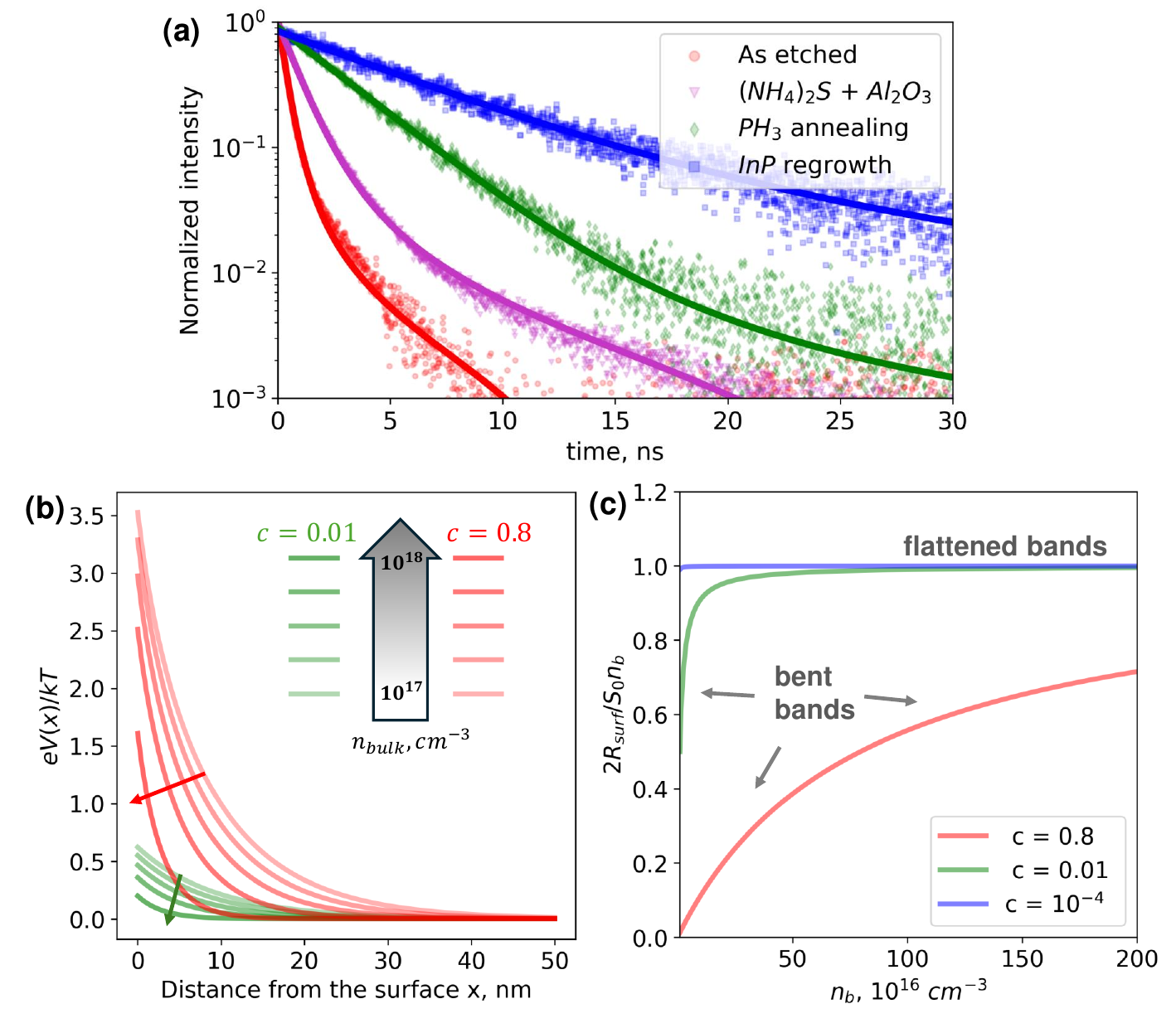} 
  \caption{\figlab{a}~Normalized TRPL traces (dots) and corresponding fit lines using \eqnref{3} and \eqnref{8} for $\SI{200}{\nano\meter}$-wide ridges with different surface treatments,
  \figlab{b}~model curves for band bending energy $eV_s$, \figlab{c}~model curves for surface recombination rate as the function of charge carrier density $n_b$.}
  \label{fig:3}
\end{figure*}

In the following, we use the model discussed above to explain carrier lifetime variation qualitatively. Solving \eqnref{4} for the case of prevailing surface recombination with $R_{\rm surf}$ from \eqnref{7} and using the simplification $n_b \approx p_b \approx N$, we obtain:

\begin{equation}
     \frac{N}{N_0} = \frac{c}{W[c  \exp(S_0 t/2 + c)]}, 
    \label{eq:8}
\end{equation}
where $W[x]$ is the Lambert $W$-function. 

\subfigrefL{3}{a} shows the TRPL curve acquired from $\SI{200}{\nano\meter}$ wide ridges and corresponding fits with \eqnref{3} and \eqnref{8}.
The fits give $c = 0.8 \pm 0.4$ for as etched structures, $c = \SI{0.6(2)}{}$ for the case of (NH$_4$)$_2$S + Al$_2$O$_3$ treatment, and $c = \SI{0.010(5)}{}$ for PH$_3$-annealed structures, which in the latter case may witness the reduction of the charge density on the sidewalls.
In the case of InP regrowth, the fit gives almost zero value of $c$ far below the fitting error, corresponding to the effective reduction of the surface charges and recombination rate $R_{\rm surf} = S_0  n_b /2$ linear with the carrier density. 

The model discussed above also explains the mechanism behind the non-linear increase of the surface recombination with the carrier density in the presence of surface charges. 
\subfigrefL{3}{b} illustrates the decrease in the band bending $eV_s$ with the increase of the carrier density. In turn, band flattening reduces the carrier depletion and thus restores the carrier density at the surface.
Higher carrier density at the surface leads to more intensive surface recombination (or shorter lifetime) at higher levels of TRPL intensity (higher levels of charge carrier concentration).

To illustrate the non-linear increase in surface recombination rate with the rise in charge carrier concentration, \subfigref{3}{c} shows the ratio between surface recombination rate $R_{\rm surf}$ and its value in the absence of surface charges, $R_{\rm surf} = S_0  n_b/2$, as a function of carrier density for the densities of surface charges corresponding to $c = 0.01$ and $c = 0.8$.
The ratio $2 R_{\rm surf}/ S_0  n_b $ in \subfigref{3}{c} stabilizes for the highest carrier densities due to screening of surface charges and flattening of the bands. Thus, for a fair comparison of different passivation techniques with the approach of \eqnref{1}, one should use recombination rates at higher intensities. This recombination rate can be found as the slopes of the TRPL curve in the log scale at the maximum intensity.

\begin{figure}[htb]
  \centering
  \includegraphics[width=\textwidth]{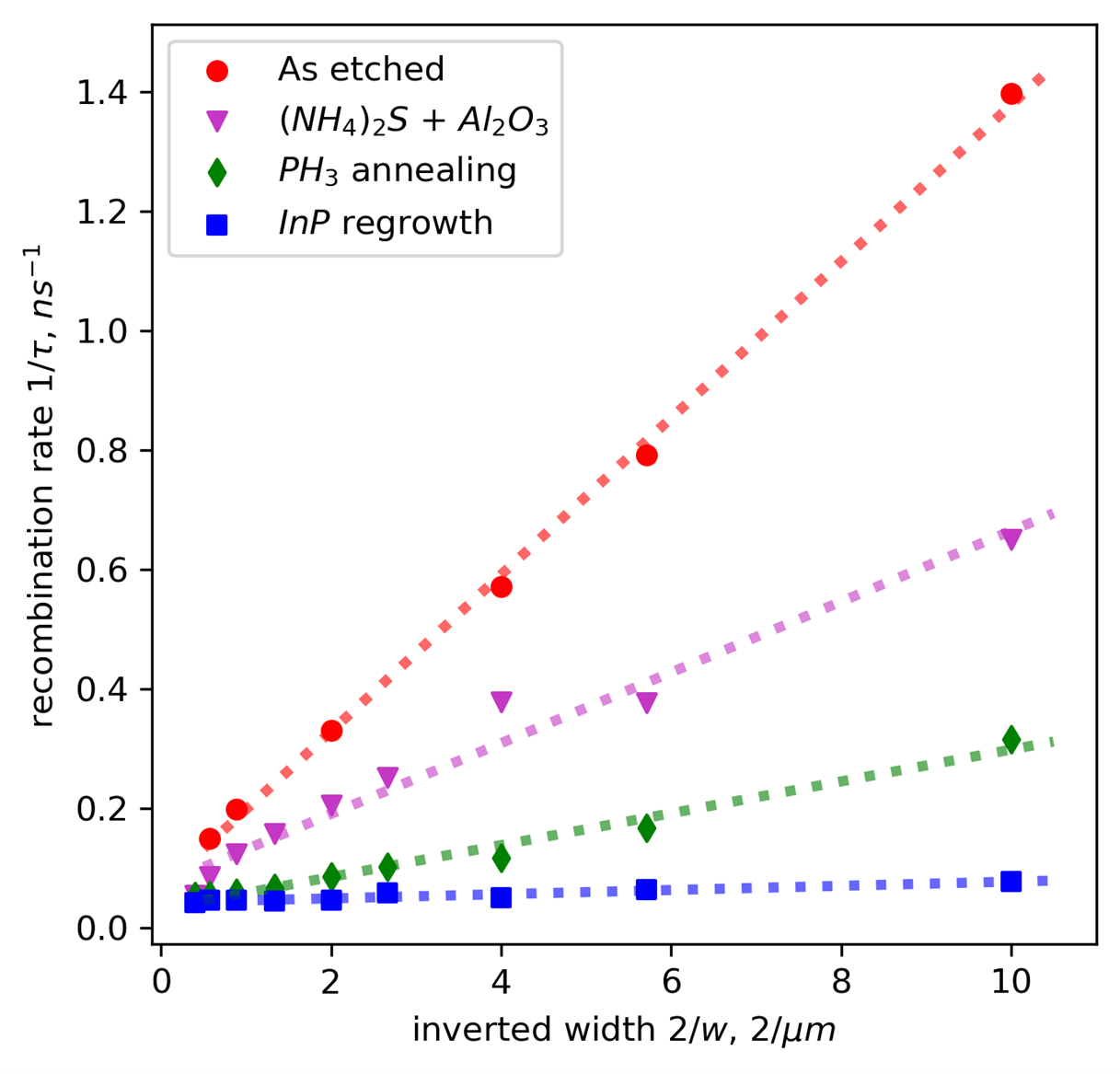} 
  \caption{Recombination rate $1/\tau$ for the investigated samples as a function of inverted width of the nanoridge $2/w$.}
  \label{fig:4}
\end{figure}

\figrefL{4} shows the obtained dependencies of the recombination rates as the function of the inverse ridge width $2/w$ for samples with different surface treatments.
The slopes of the obtained linear dependencies give the corresponding surface recombination velocities $s$, summarized in \tabref{surf-recomb-velocity}.

\begin{table}\centering
\ra{1.15}
\caption{Fitted surface recombination velocities for investigated samples.}
\label{tab:surf-recomb-velocity}
\begin{tabular}{cc}\toprule
Passivation method & Surface recombination velocity $s$ \\\midrule
As etched & $\SI{1.5E4}{\centi\meter\per\second}$ \\
(NH$_4$)$_2$S + Al$_2$O$_3$ & $\SI{5E3}{\centi\meter\per\second}$\\
PH$_3$ annealing & $\SI{2E3}{\centi\meter\per\second}$\\
InP regrowth & $\SI{5E2}{\centi\meter\per\second}$\\
\bottomrule
\end{tabular}
\end{table}

\section*{Conclusion}
In this work, we report on a new approach to passivate nanostructures with III-As(P) material fabricated by the dry etching process.
To heal the defects formed during dry etching and subsequent native oxide formation on the etched sidewalls, we anneal the nanostructures under phosphine ambient and compare the results with different passivation methods previously known in the literature.
Our findings demonstrate an order of magnitude reduction of surface recombination rate from $\SI{1.5E4}{\centi\meter\per\second}$ in non-passivated structures to $\SI{2E3}{\centi\meter\per\second}$ in the case of PH$_3$ annealing.
The latter value is also lower than $\SI{5E3}{\centi\meter\per\second}$ obtained for samples with widely used passivation by (NH$_4$)$_2$S wet treatment followed by ALD of Al$_2$O$_3$.
However, it is still higher than the recombination rate of $\SI{5E2}{\centi\meter\per\second}$ estimated for the complete regrowth of the ridges with InP.

Our TRPL studies reveal the surface-related increase in carrier lifetime at lower intensities (and thus at lower charge carrier densities) for non-passivated and (NH$_4$)$_2$S + Al$_2$O$_3$-treated samples but less pronounced in the samples treated by PH$_3$ annealing.
We explain this observation by less effective screening of the surface fields, which depletes the carriers in the vicinity of the sidewalls.
The depletion, in turn, reduces the surface recombination rate, which increases the carrier lifetime in the case of surface-defined recombination in narrow ridges.
The successful implementation of this surface passivation method in nanolaser structures based on extreme dielectric confinement cavities highlights its potential.
The complicated shape of the nanolaser cavity was preserved during the in-situ MOVPE annealing in PH$_3$ ambient, while the passivation allowed achieving the stable continuous-wave laser emission in the telecom C-band at room temperature~\cite{xiong2024room, Xiong2024}.\\

\appendix
\section{Band bending at the surface}\label{SI:Band-bending}

We assume the validity of the quasi-Fermi level approximation, which implies exponential dependences for $p=p_b\exp(-v(x))$ and $n_s=n_b\exp(v(x))$ with $v(x)\equiv eV(x)/kT$ and $n_b$ and $p_b$, denoting the carrier concentrations far from the surface. At the first step, the surface potential $V(x)$ within the space-charge regions can be found from the Poisson equation~\cite{monch2013semiconductor, luth2013surfaces}:

\begin{multline}\label{eq:s1}
\frac{\dd[2]{V(x, n_b, p_b, Q_s)}}{\dd{x}^2} = \\
-\frac{e}{\epsilon \epsilon_0}\left(n_b(e^{eV/kT}-1) -(p_b(e^{-eV/kT}-1)\right),
\end{multline}
where $x$ is the distance from the sidewall surface. In the following, we assume $n_b = p_b = N$.
By definition, the surface field $E$ is the derivative of the potential $V$, $E = -\dd{V}/\dd{x}$, and thus $\dd{E^2}/\dd{V} = 2 \dd[2]{V}/\dd{x^2}$.
Inserting the latter into \eqnref{s1} and integrating over $V$ with the initial condition $E(V=0) = 0$, one obtains:

\begin{multline}
E^2(V) = 4\frac{kT}{\epsilon} N_0 \left(\cosh(eV/kT)-1\right) = \\    
8\frac{kT}{\epsilon} N \left(\sinh^2\left(\frac{eV}{2kT}\right)\right).\label{eq:s2}
\end{multline}

Using the notation $v=eV/kT$, this yields:

\begin{equation}
\frac{\dd{v}}{\dd{x}} = -2\sqrt{2\frac{e^2}{\epsilon \epsilon_0 kT} N} \sinh(v/2).\label{eq:s3}
\end{equation}

Integration of \eqnref{s3} by parts with the boundary condition $v(x=0)=v_s$ leads to:
\begin{equation}
\tanh(\frac{v}{4})=\tanh\left(\frac{v_s}{4}\right)\exp\left(-x\sqrt{2\frac{e^2}{\epsilon \epsilon_0 kT}N}\right).\label{eq:s4}
\end{equation}

To estimate $n_s$ and $v_s$, we follow the well-known approach based on the balance between the surface charge density $Q_s$ and the charge density within the space-charge regions $Q_{sc}=-Q_s$. Assuming surface defects to be fully ionized, $Q_s$ does not change with carrier concentration ($Q_s = $ \textit{const}). The charge density in the space-charge region $Q_{sc}$ can be expressed in terms of the surface field $E_s$ as $Q_{sc} = \epsilon \epsilon_0 E_s$.
Therefore, \eqnref{s2} leads to:
\begin{equation}
    -Q_s = 2\sqrt{2\epsilon \epsilon_0 kT N} \sinh(v_s/2).\label{eq:s5}
\end{equation}

Denoting the density of surface charge as $N_s=|Q_s|/e$, from \eqnref{s5} the surface potential $v_s$ can be found as a function of $N_s$ and $N$:

\begin{equation}
v_s(N_s, N) = 2\arcsinh\left(\frac{1}{2\sqrt{2}}\sqrt{\frac{e^2}{\epsilon \epsilon_0 kT}} \frac{N_s}{\sqrt{N}}\right).\label{eq:s6}
\end{equation}

Then, knowing $v_s$, the carrier densities at the surface are given by $n_s = N \exp(v_s)$, and $p_s = N \exp(-v_s)$.

\section{Surface recombination rate}\label{SI:Surface-recombination-rate}

The Shockley-Read-Hall model gives the following expression for the surface recombination rate $R_{\rm surf}$:

\begin{equation}
    R_{\rm surf}=\frac{p_s n_s-n_i^2}{S_p^{-1} (n_s + n_t)  + S_n^{-1}  (p_s + p_t)},\label{eq:s7}
\end{equation}
where $p_s$ and $n_s$ are hole and electron densities in the vicinity of the sidewall surfaces, $n_t$  and $p_t$ are corresponding trap densities, and $S_p$ and $S_n$ are the corresponding surface recombination rates. We take $S_p = S_n = S_0$ for simplicity in the following. 
Neglecting trap density and intrinsic carrier densities compared to photogenerated carrier densities, we obtain:
\begin{equation}
    R_{\rm surf} = S_0 \frac{p_b n_b}{n_b\exp(v_s)+p_b\exp(-v_s)} = \frac{S_0 N}{2\cosh(v_s)}.\label{s7}
\end{equation}

Using \eqnref{s6} and the property $\cosh(x) = 1+2\sinh^2(x/2)$ we have:
\begin{equation}
    R_{\rm surf} = \frac{1}{2} {S_0 N}/\left(1+\frac{e^2}{4\epsilon \epsilon_0 kT} \frac{N_s^2}{N}\right).\label{eq:s9}
\end{equation}

In case of non-negligible $\frac{e^2}{4\epsilon \epsilon_0 kT}\frac{N_s^2}{N}$, the obtained expression for $R_{\rm surf}$ is non-linear with the photogenerated carrier densities.

Next, we can use the obtained expression for $R_{\rm surf}$ to estimate the carrier dynamics. In the case of dominant surface recombination $\dd{N}/\dd{t} \approx -R_{\rm surf}$, which leads to the expression in \eqnref{8}:

\begin{equation}
    N = N_0 \frac{c}{W[c \exp(S_0 t/2 + c)]}, \label{eq:s10}
\end{equation}
where $W[x]$ is the Lambert $W$-function, and we introduce the coefficient $c=\frac{e^2}{4\epsilon \epsilon_0 kT}\frac{N_s^2}{N_0}$. 
The Lambert function has an asymptote $W[x \to 0] \to x$, so an exponential dependence $n_b \approx \exp(-S_0 t/2)$ is expected at small $c$.\\

\noindent{\bf \normalsize Data Availability}\\
Data underlying the results presented in this paper may be obtained from the authors upon reasonable request.\\

%\newpage
%\bibliography{thebib}

\vspace*{2mm}
\noindent{\bf \normalsize Acknowledgment}\\
The authors acknowledge the financial support from the Danish National Research Foundation through NanoPhoton -- Center for Nanophotonics, grant number DNRF147, and bigQ -- Center for Macroscopic Quantum States, grant number DNRF142.
Y.\,B. thanks all the colleagues at DTU and Wroc\l{}aw University of Science and Technology for a fruitful collaboration.\\

\noindent{\bf \normalsize Funding}\\
Danish National Research Foundation: DNRF147, Y.\,B., P.\,H., K.\,Y., and E.\,S.; DNRF142, A. H.;\\

\noindent{\bf \normalsize Author Contributions}\\
Y.\,B. carried out epitaxy, nanofabrication, and surface state passivation experiments of ridge samples.
A.\,S. and K.\,Y. provided buried heterostructure samples.
Y.\,B., P.\,H., J.\,M.\,S., E.\,Z.-O., under the lead of M.\,S. performed optical measurements.
Y.\,B., A.\,H., P.\,H., M.\,S., E.\,S. interpreted the data.
Y.\,B., with the contribution of A.\,H. developed the model.
Y.\,B., A.\,H., P.\,H., and E.\,S. wrote the manuscript with input from all authors.
E.\,S. conceived and supervised the project.\\

\noindent{\bf \normalsize Disclosures}\\
The authors declare no conflicts of interest.\\

\end{document}